\documentclass[11pt]{article}
\usepackage[total={6in, 9.1in}]{geometry}
\usepackage{url}

\usepackage{amsthm}
\theoremstyle{definition}
\newtheorem{definition}{Definition}[section]
\newtheorem{lemma}{Lemma}[section]

\usepackage{afterpage}

\usepackage{amsmath}
\usepackage[nameinlink, capitalise]{cleveref}
\usepackage{todonotes}
\presetkeys%
    {todonotes}%
    {inline}{}

\usepackage[english]{babel}    % allows underscore in label https://tex.stackexchange.com/a/315666
\usepackage[strings]{underscore}   % fixes problem with underscore in DOI https://tex.stackexchange.com/a/315666

\usepackage{algorithm}
\usepackage{algpseudocode}
\newcounter{protocol}
\makeatletter
\newenvironment{protocol}[1][htb]{%
  \let\c@algorithm\c@protocol
  \renewcommand{\ALG@name}{Protocol}% Update algorithm name
  \begin{algorithm}[#1]%
  }{\end{algorithm}
}
\makeatother
\crefname{prot}{Protocol}{Protocols}
\Crefname{prot}{Protocol}{Protocols}

\usepackage{graphicx}
\begin{document}
% or Essential Properties?
\title{Fundamental Properties of the Layer Below \\ a Payment Channel Network\\ {\it Extended Version}
}
\author{Matthias Grundmann \and Hannes Hartenstein}
\date{%
    Karlsruhe Institute of Technology, Karlsruhe, Germany\\%
    firstname.lastname@kit.edu%
}

\maketitle              % typeset the header of the contribution

\makeatletter
\newcommand{\modelName}{\textsc{RFL} model}
\newcommand{\utxo}[2]{\left(#1\;\middle|\;#2\right)}
\newcommand{\newOwners}[1]{\Omega^\mathrm{R}_{#1}}
\newcommand{\oldOwners}[1]{\Omega^\mathrm{S}_{#1}}
\newcommand{\owners}[1]{\Omega_{#1}}
\newcommand{\signatureA}{\sigma_\mathrm{A}}
\newcommand{\signatureB}{\sigma_\mathrm{B}}
\newcommand{\channelIdentifier}{C_\mathrm{AB}}
\newcommand{\commitmentTX}[2]{t_{\mathrm{C}{#1}\mathrm{#2}}}
\newcommand{\timeoutTX}[2]{t_{\mathrm{T}{#1}\mathrm{#2}}}
\newcommand{\successTX}[2]{t_{\mathrm{S}{#1}\mathrm{#2}}}
\newcommand{\fundingTX}{t_\mathrm{F}}
\newcommand{\balance}[1]{\mathcal{B}^\mathrm{#1}}
\newcommand{\HTLCset}[2]{\mathcal{H}^\mathrm{#1}_{#2}}
\newcommand{\HTLCsetOut}[2]{\overrightarrow{\mathcal{H}}^\mathrm{#1}_{#2}}
\newcommand{\HTLCsetIn}[2]{\overleftarrow{\mathcal{H}}^\mathrm{#1}_{#2}}
\newcommand{\deltaConfirm}{\Delta l_\mathrm{conf}}
\newcommand{\deltaSync}{\Delta l_\mathrm{sync}}
\newcommand{\deltaTxComm}{\Delta t_\mathrm{comm}}
\newcommand{\timeTxHTLC}[1]{T^\mathrm{HTLC}_{#1}}
\newcommand{\timeTxHTLCMax}{T^\mathrm{HTLC}_\mathrm{max}}
\newcommand{\timeNow}{T_\mathrm{now}}
\newcommand{\deltaForward}{\Delta_\mathrm{forw}}    % LN: cltv_expiry_delta
\newcommand{\deltaUserCheck}{\Delta_\mathrm{user}}
\newcommand{\channelId}{\mathtt{channel\_id}}
\newcommand{\channelIds}{\mathtt{channel\_ids}}
\newcommand{\tmpChannelId}{\mathtt{tmp\_id}}
\newcommand{\tmpChannelIds}[1]{\mathtt{tmp\_ids_{#1}}}
\newcommand{\channelIdMap}[1]{\mathtt{channel}^{#1}}
\newcommand{\channelOpen}[1]{\mathtt{opened}^{#1}}

\newcommand{\figureSubnum}[1]{$\langle #1 \rangle$}

\makeatother

\begin{abstract}
Payment channel networks are a highly discussed approach for improving scalability of cryptocurrencies such as Bitcoin.
As they allow processing transactions off-chain, payment channel networks are referred to as second layer technology, while the blockchain is the first layer.
We uncouple payment channel networks from blockchains and look at them as first-class citizens.
This brings up the question what model payment channel networks require as first layer.
In response, we formalize a model (called \modelName{}) for a first layer below a payment channel network.
While transactions are globally made available by a blockchain, the \modelName{} only provides the reduced property that a transaction is delivered to the users being affected by a transaction.
We show that the reduced model's properties still suffice to implement payment channels.
By showing that the \modelName{} can not only be instantiated by the Bitcoin blockchain but also by trusted third parties like banks, we show that the reduction widens the design space for the first layer.
Further, we show that the stronger property provided by blockchains allows for optimizations that can be used to reduce the time for locking collateral during payments over multiple hops in a payment channel network.
\end{abstract}

\section{Introduction}

Payment channel networks (PCNs) became popular as an approach for improving the scalability of blockchain based cryptocurrencies such as Bitcoin \cite{nakamoto_bitcoin:_2008}.
While Bitcoin scales well in the amount of coins that can be transferred by a transaction, it can only process a limited number of transactions per second.
PCNs, e.g., the Lightning Network \cite{poon_bitcoin_2016} for Bitcoin, perform transactions off-chain in a second layer and they do not require global consensus for every transaction as long as all participants are honest.

PCNs have mostly been looked at as second layer on top of a blockchain.
While the idea to use banks as a first layer has already been mentioned in 2015 by Tremback and Hess \cite{tremback_universal_2015}, to the best of our knowledge it has not been analyzed which properties PCNs require for the layer below.
We look at PCNs as first-class citizen independent from the specific first layer and analyze whether PCNs can be used on top of models that are different from a blockchain (with the term blockchain, we refer to a public permissionless blockchain such as Bitcoin).
As an affirmative answer, we present a reduced model for the first layer which we call \modelName{} (Reduced First Layer) whose properties give reduced guarantees compared to a blockchain but yet suffice to implement a protocol for a PCN on top.
The reduced property of the \modelName{} in comparison to a blockchain is that a blockchain delivers each transaction to all peers participating in the network while in the \modelName{} a transaction is only guaranteed to be visible for a transaction's affected users.
A transaction's affected users are the users who receive the transferred coins and the users who were able to spend the same coins that the transaction transfers.
To show that a PCN can still securely be implemented using the \modelName{} as underlying first layer, we show the following property:
A payment channel between two users can be closed within a given time so that each user $u$ receives at least their correct balance if $u$ is honest and checks the first layer regularly for new transactions.
We will refer to this property as the security property and formally define it.
We present a slightly simplified version of the Lightning Network's protocol and we prove this security property for this protocol when it is executed on a first layer that implements the properties defined by the \modelName{}: liveness, affected user synchrony, persistence, and transaction validity.

Our analysis of the relationship between the \modelName{} and a blockchain shows that a blockchain instantiates the \modelName{} under typical assumptions for blockchains such as Bitcoin.
Having shown that a PCN can be implemented on a reduced model compared to a blockchain, we show that the \modelName{} can not only be instantiated by blockchains but there is a wider design space for the first layer, e.g., using trusted third parties like banks.
Implementation of PCNs on different first layers allows for a range of design opportunities, e.g., with respect to trust, privacy, liquidity, online requirements, and currencies.
Having seen the advantages of a model for the first layer that is reduced compared to a blockchain, we also look at the advantages of using a blockchain that guarantees more than the \modelName{}:
A blockchain's property to deliver a transaction to everyone can be used to optimize payments in PCNs so that collateral is locked for a shorter amount of time \cite{miller_sprites_2017}.
We show that this requires stronger assumptions than the basic construction of PCNs.

To summarize, we present the following contributions in this paper:
\begin{itemize}
	\item A formal model (\modelName) of a reduced first layer compared to a blockchain.
	\item We show that a protocol for PCNs is secure in the proposed model.
	\item The \modelName{} can be instantiated using blockchains and other architectures allowing for a range of design options.
	\item An analysis showing that a blockchain's additional properties can enable optimizations for payments in a PCN.
\end{itemize}

In the following section, we introduce the fundamentals of payment channels and PCNs.
In \cref{sec-model-first-layer}, we present the \modelName{}.
After presenting a protocol for PCNs in \cref{sec-protocol-pcn}, we show in \cref{sec-security} that this protocol fulfills the security property if it is used on a first layer that instantiates the \modelName.
\Cref{sec-instances} presents and compares instances of the \modelName{} using a blockchain and trusted banks.
In \cref{sec-optimizations}, we show how the stronger properties fulfilled by a blockchain can be used to optimize payments in a PCN.
After putting our work in the context of related work in \cref{sec-related-work}, we conclude in \cref{sec-conclusion}.

\section{Fundamentals}
\label{sec-fundamentals}

By introducing the Lightning Network \cite{poon_bitcoin_2016}, Poon and Dryja published the idea of building a network of payment channels and using it as a second layer for payments which happen off-chain instead of on-chain.
In this section, we give a simplified overview of the functionality and explain the protocol more formally and in more detail in \cref{sec-protocol-pcn}.

A payment channel has two participants who create the channel by locking coins in a 2-of-2 multisig address using a funding transaction.
These coins can be spent only by transactions that are signed by both of the channel's participants.
During the lifetime of the channel, both participants have commitment transactions signed by the other party which distribute the channel's funds to the participants according to the current balance.
Payments can be made between the participants by creating a new commitment transaction with an updated balance.
With each channel update, the participants exchange revocation keys which can be used to spend a dishonest participants' balance in case the dishonest participant publishes an outdated commitment transaction.
The channel can be closed either by publishing the last commitment transaction or by creating a specific closing transaction.

Opening a payment channel for just one transaction is not worth the cost of opening and closing a payment channel which requires the payment of transaction fees for two transactions on-chain.
To remove the necessity of opening a payment channel for every new payment partner, it is possible to route payments over other participants in a network of payment channels.
To securely perform a payment over multiple hops, it is required that intermediate nodes can be assured that their outgoing payments succeed iff their incoming payment succeeds.
The Lightning Network uses Hash Timelocked Contracts (HTLCs) to ensure this atomicity of payments.
For a payment from Alice to Charlie of $c$ coins over the intermediate Bob,
Alice creates a secret $x$ and uses a hash function $H$ to get $y = H(x)$.
Alice then updates her channel with Bob to a state that has a conditional output that gives $c$ coins to Bob under the condition that Bob can provide a preimage for $y$ before the time given by $T_\mathrm{AB}$ has passed.
Bob updates his channel with Charlie accordingly with a conditional output that gives Charlie $c$ coins\footnote{For simplicity, we leave out the fee that Bob subtracts for forwarding the payment.} before $T_\mathrm{BC} < T_\mathrm{AB}$ time has passed.
At this time, Bob has an \textit{incoming} HTLC in his channel with Alice and an \textit{outgoing} HTLC in his channel with Charlie.
For Charlie to receive the payment, Alice sends the secret $x$ directly to Charlie.
Charlie sends $x$ to Bob to redeem the conditional output who forwards it to Alice to redeem the conditional output in their channel.
When Bob receives the secret $x$ from Charlie, he has time to forward it to Alice to redeem his incoming HTLC as well because $T_\mathrm{BC} < T_\mathrm{AB}$.
In case Bob's outgoing HTLC to Charlie is not redeemed in time, he will not be able to redeem his incoming HTLC.
Using this construction, atomicity is ensured for payments using HTLCs.

\section{RFL Model of First Layer} % \modelname not used here for title case
\label{sec-model-first-layer}

In this section, we present the \modelName{}, a model for the first layer that guarantees a reduced set of properties compared to a blockchain.
The main difference of the \modelName{} to a blockchain is that, when using a blockchain, a transaction is delivered to all users.
In this section, we define the \textit{affected user synchrony} property which only requires the first layer to deliver a transaction to the users being affected by a transaction.

In the \modelName{}, we have a set of users $U$ who can create transactions.
Each user $u \in U$ has an asymmetric key pair with private key $s_u$ and the public key $p_u$.
A transaction $t$ consists of an amount of coins and is associated with a set of \textit{receivers} $\newOwners{t} \subseteq U$ that the coins are transferred to.
The set of users who were able to spend the coins that are spent by $t$ is referred to as the \textit{potential senders} $\oldOwners{t} \subseteq U$.
We denote as \textit{affected users} of the transaction $t$ the set $\owners{t} = \oldOwners{t} \cup \newOwners{t}$.
For a transaction to be valid, it needs to fulfill conditions depending on the coins that are spent (e.g., a signature using a specified key or some timeout having passed).
In general, it is possible that not all potential senders need to sign a transaction and thus some potential senders and receivers might not have seen the transaction before it has been published on the first layer.

We model the first layer as a (logical) single party $\mathcal{L}$ that is connected to all other parties via secure and reliable channels.
Users can send transactions $t$ to the first layer $\mathcal{L}$.
We refer to this action as \textit{publishing} $t$.
$\mathcal{L}$ can send a confirmation that the transaction $t$ has been executed, i.e. the coins have been transferred, and users can query $\mathcal{L}$ whether a transaction has been confirmed.
The time a user has to wait for the confirmation is determined by the liveness property parametrized below by the waiting time $\deltaConfirm$.
The first layer $\mathcal{L}$ can send transactions from other users and confirmations to users.
Users can check the first layer $\mathcal{L}$ for the confirmation of a transaction.
Whether the first layer's response about the confirmation is consistent among different users, is determined by the affected user synchrony property below parametrized by the time $\deltaSync$.

The first layer $\mathcal{L}$ has to have these essential properties:
\begin{itemize}
  \item \textit{Liveness}: If a user sends a valid transaction $t$ to the first layer $\mathcal{L}$, then $\mathcal{L}$ will confirm the transaction after at most $\deltaConfirm$.
  \item \textit{Affected User Synchrony}: If a user has received a confirmation from $\mathcal{L}$ for a transaction $t$ with affected users $\owners{t} $, then $\mathcal{L}$ makes $t$ and the confirmation visible to all affected users $u \in \owners{t} $ within at most $\deltaSync$.
  \item \textit{Persistence}: If a user has received a confirmation for transaction $t$ from $\mathcal{L}$, then $\mathcal{L}$ will always report $t$ as confirmed.
  \item \textit{Transaction Validity}: A transaction $t$ will only be confirmed by $\mathcal{L}$ if $t$ is valid.
\end{itemize}

Note that instead of the affected user synchrony, a blockchain implements an unrestricted synchrony property:
If a transaction is confirmed by the blockchain, the transaction will be seen as confirmed by all users within a given time.
We will look deeper into the relationship between the \modelName{} and blockchains in \cref{sec-instances}.

For the protocol in the following section, we use the UTXO (unspent transaction output) model for transactions which is also used by Bitcoin.
The protocol could also be transformed to an account model (e.g., as used in Ethereum).
In the UTXO model, a transaction consists of multiple inputs and multiple outputs.
Each input spends an output of a previous transaction.
An output specifies a condition that an input needs to meet to spend the output.
An example for a condition is the signature of a public key that is specified in the output and the signature needs to be provided in the input.
If a transaction is processed by the first layer, the first layer checks whether the transaction is \textit{valid}.
For a transaction to be valid, all inputs have to spend transaction outputs that are unspent and the conditions of the UTXOs need to be met.
For a PCN as defined later, we require the following types of conditions: signature corresponding to a given public key, preimage for a given image of a hash function, time since confirmation of UTXO spent, and combinations thereof using logical operators OR and AND.
The set of receivers $\newOwners{t}$ contains all users whose signature is required by at least one condition to spend an output of $t$.
Analogously, the set of potential senders $\oldOwners{t}$ of a transaction $t$ is comprised of all users whose signature is required by at least one condition to spend an output that is spent by $t$.

Having defined the \modelName{}, we present a protocol for PCNs in the following section for which we will show that it can securely be implemented using the \modelName{}.

\section{Protocol for a Payment Channel Network}
\label{sec-protocol-pcn}

Before we can prove in \cref{sec-security} that the \modelName{} suffices as a first layer for a secure implementation of a PCN protocol, we need to specify a protocol for a PCN that can be used in the proof.
To this end, we present in this section a protocol for a PCN that is a slightly simplified version of the \textbf{Lightning Network's protocol} specified on Github\footnote{\scriptsize{\url{https://github.com/lightningnetwork/lightning-rfc/blob/master/00-introduction.md}}}.
The protocol assumes a first layer $\mathcal{L}$ that instantiates the \modelName{} presented in the previous section.
While explaining the protocol for PCNs, we show the relation between the \modelName{} and the protocol by making explicit where the properties of the \modelName{} are used.
The current specification of the Lightning Network's protocol is only for single-funded channels, i.e. a channel is created by putting funds of only one participant in the channel.
Payments are executed using HTLCs even if they are direct payments over just one channel.

We present the protocol for PCNs by splitting it into four smaller protocols:
A payment channel is opened using \cref{protocol:overviewOpenChannel}, payments are performed using \cref{protocol:overviewPaymentOverIntermediaries}, and channels are closed using \cref{protocol:overviewCloseChannel}.
As a subprotocol of \cref{protocol:overviewPaymentOverIntermediaries}, \cref{protocol:overviewUpdateChannel} shows how a channel between two parties is updated to a new state when an HTLC is added or removed.
The protocols use Alice and Bob as names for two users participating in the protocol.
These names are to be read as variables and the user called Alice in one protocol can be a different user in another protocol.
We count the states of a channel using $n \geq 1$ and denote the commitment transaction held by Alice for state $n$ as $\commitmentTX{n}{A}$ ($\commitmentTX{n}{B}$ for Bob).
For revocation, Alice creates a new revocation key pair with the secret key $s_{\mathrm{R}n\mathrm{A}}$ and the public key $p_{\mathrm{R}n\mathrm{A}}$ for each commitment transaction $\commitmentTX{n}{A}$.
Each output of Alice's commitment transaction $\commitmentTX{n}{A}$ is spendable using the revocation key $s_{\mathrm{R}n\mathrm{A}}$ and Bob's public key, i.e. Bob can spend all outputs of $\commitmentTX{n}{A}$ if Alice publishes $\commitmentTX{n}{A}$ after she has revoked $\commitmentTX{n}{A}$ by sending $s_{\mathrm{R}n\mathrm{A}}$ to Bob.

The protocol to open a channel is shown in \cref{protocol:overviewOpenChannel}.
Alice chooses one of her unspent outputs $o_\mathrm{A}$ of amount $a$ that she wants to fund the channel with.
Alice creates a revocation key pair with the public key $p_\mathrm{R1A}$ and the secret key $s_\mathrm{R1A}$ and sends her public keys  $p_\mathrm{A}$ and $p_\mathrm{R1A}$ and her funding amount $a$ to Bob.
Bob replies with his public key $p_\mathrm{B}$.
Alice creates the funding transaction $\fundingTX$ and Bob's version of the initial commitment transaction $\commitmentTX{1}{B}$.
As the channel is single-funded, the funding transaction $\fundingTX$ sends $a$ coins from Alice's unspent output $o_\mathrm{A}$ to an output that can only be spent jointly by Alice and Bob.
The initial commitment transaction sends all funds of the channel back to Alice.
The general construction of commitment transactions will be explained in the following paragraph.
Alice signs $\commitmentTX{1}{B}$ and sends her signature of $\commitmentTX{1}{B}$ and the id of the funding transaction to Bob.
Bob needs the id of the funding transaction to create Alice's version of the commitment transaction $\commitmentTX{1}{A}$ because the commitment transaction spends the funding transaction's output.
Bob creates and signs $\commitmentTX{1}{A}$ and sends the signature to Alice.
Having the signature for the initial commitment transaction, Alice can securely open the channel by signing the funding transaction $\fundingTX$ and publishing $\fundingTX$ by sending it to the first layer $\mathcal{L}$. 
According to the liveness property, the first layer will confirm the funding transaction $\fundingTX$ within $\deltaConfirm$ and Alice and Bob will see the confirmation within $\deltaSync$ according to the affected user synchrony property of the first layer $\mathcal{L}$.
In preparation for future updates of the channel, Alice and Bob create new revocation key pairs and exchange their public keys once they receive the confirmation for $\fundingTX$ from $\mathcal{L}$. % longer: the first layer $\mathcal{L}$

\begin{protocol}[tp]
  \caption{Open a payment channel between Alice and Bob}
  \label[prot]{protocol:overviewOpenChannel}
  \textit{Inputs.} Alice: amount $a$ to fund the channel with.
  \\
  \textit{Goal.} A payment channel between Alice and Bob has been opened. The funding transaction $\fundingTX$ has been sent to $\mathcal{L}$ and Alice (resp. Bob) has $\commitmentTX{1}{A}$ ($\commitmentTX{1}{B}$).
  \begin{algorithmic}[1]
    \State Alice chooses one of her unspent outputs $o_\mathrm{A}$ of amount $a$.
    \State Alice creates a key pair with the public key $p_\mathrm{R1A}$ and the secret key $s_\mathrm{R1A}$.
    \State Alice sends her public key $p_\mathrm{A}$, her funding amount $a$, the public key $p_\mathrm{R1A}$ to Bob. %(\textsc{open channel})
    \State Bob sends his public key $p_\mathrm{B}$ to Alice. %(\textsc{accept channel})
    \State Alice creates $\commitmentTX{1}{B}$ and the funding transaction $\fundingTX$.
    \State Alice sends her signature of $\commitmentTX{1}{B}$ and the id of $\fundingTX$ to Bob. %(\textsc{funding created})
    \State Bob creates Alice's version of the initial commitment transaction $\commitmentTX{1}{A}$.
    \State Bob sends his signature of $\commitmentTX{1}{A}$ to Alice. %(\textsc{funding signed})
    \State Alice signs $\fundingTX$ and sends $\fundingTX$ to $\mathcal{L}$. %(\textsc{publish funding})

    \State $\mathcal{L}$ sends a confirmation to Alice and Bob after at most $\deltaConfirm$.
    \State Alice / Bob creates a new key pair with the public key $p_\mathrm{R2A}$ / $p_\mathrm{R2B}$ and the secret key $s_\mathrm{R1A}$ / $s_\mathrm{R1B}$.
    \State Alice / Bob sends $p_\mathrm{R2A}$ / $p_\mathrm{R2B}$ to Bob / Alice.  %(\textsc{funding locked})
  \end{algorithmic}
\end{protocol}

\begin{figure}[tb]
    \includegraphics[width=0.9\textwidth]{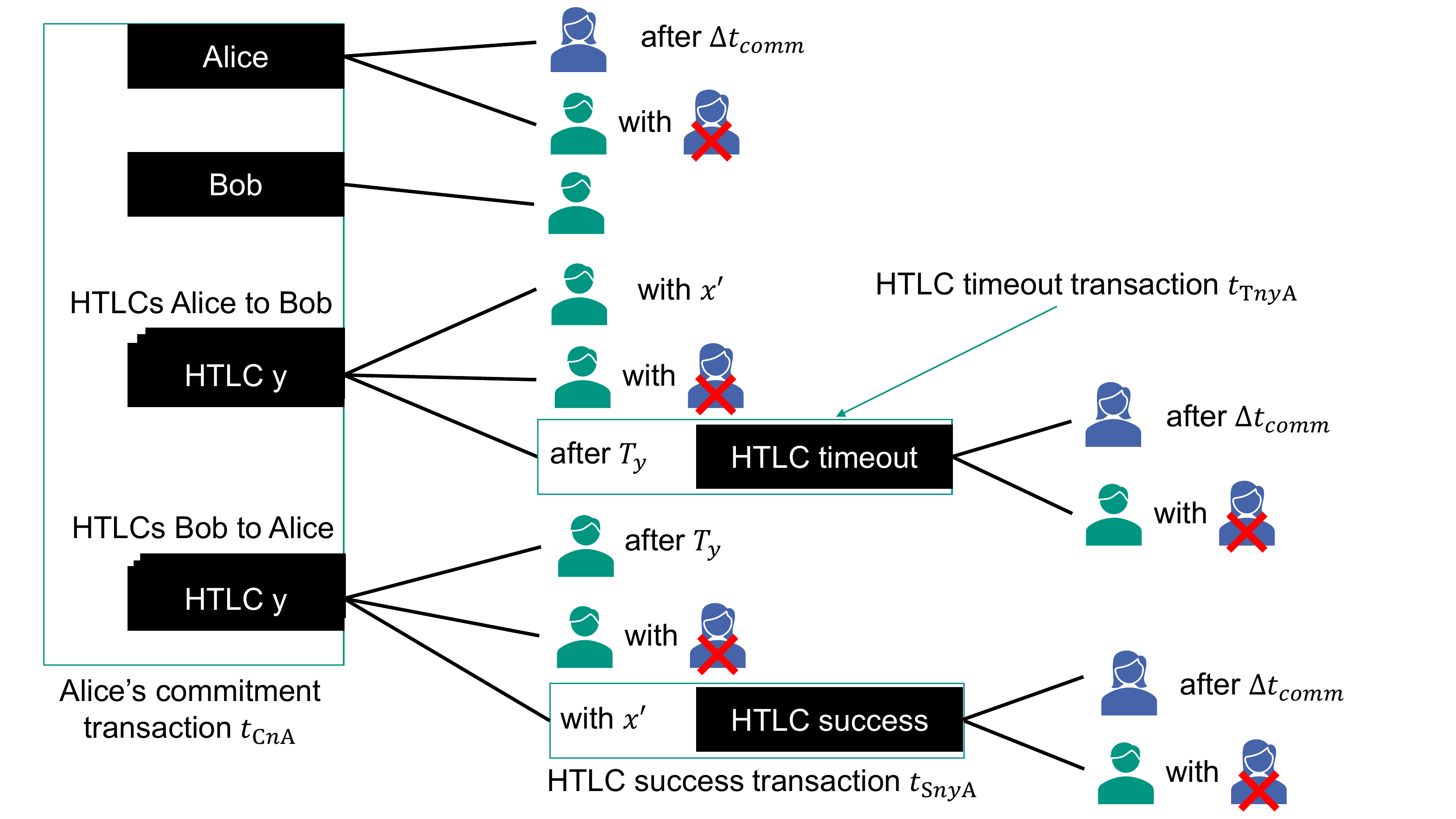}
    \caption{Possible ways for spending the outputs of Alice's commitment transaction $\commitmentTX{n}{A}$.
    Multiple lines represent the alternative ways an output can be spent.
    The blue icon indicates that Alice's secret key $s_\mathrm{A}$ is required to spend the output.
    Analogously, the green icon stands for Bob's secret key $s_\mathrm{B}$.
    The crossed-out blue icon stands for Alice's revocation key $s_{\mathrm{R}n\mathrm{A}}$.
    $x'$ is a preimage for $y$, i.e. $h(x') = y$.}
    \label{fig:commitment-tx-alice}
\end{figure}

Each commitment transaction $\commitmentTX{n}{A}$ held by Alice for state $n$ of the channel between Alice and Bob is built in the following way ($\commitmentTX{n}{B}$ held by Bob is constructed analogously):
The commitment transaction's input spends the channel's funding transaction's output which requires Alice's and Bob's signatures.
We use the term \textit{stable balance} for a user's balance that is not part of an HTLC.
The HTLC outputs of $\commitmentTX{n}{A}$ cannot directly be spent by Alice but only using the HTLC timeout transaction $\timeoutTX{ny}{A}$ and the HTLC success transaction $\successTX{ny}{A}$ which will be explained below.
Outputs spendable by Alice are locked for $\deltaTxComm$ time to give Bob time to spend the output in case Alice publishes $\commitmentTX{n}{A}$ after it has been outdated.
Alice's commitment transaction $\commitmentTX{n}{A}$ has the following outputs:
\begin{itemize}
	\item An output for Alice's stable balance that is spendable
	\begin{itemize}
	    \item by Bob using Alice's revocation key $s_{\mathrm{R}n\mathrm{A}}$ for state $n$ or
	    \item by Alice after delay $\deltaTxComm$; aka `to self delay'. % or 'to self delay'
	\end{itemize}
	\item An output for Bob's stable balance that is spendable by Bob.
	\item For each outgoing HTLC an output for the HTLC's balance that is spendable
	\begin{itemize}
		\item by Bob if he provides a preimage for a given $y$, or
		\item by Bob using Alice's revocation key for state $n$, or
		\item by the HTLC-timeout transaction $\timeoutTX{ny}{A}$ after point in time $\timeTxHTLC{y}$.
	\end{itemize}
	\item For each incoming HTLC an output for the HTLC's balance that is spendable
	\begin{itemize}
		\item by Bob after point in time $\timeTxHTLC{y}$, or
		\item by Bob using Alice's revocation key for state $n$, or
		\item by the HTLC-success transaction $\successTX{ny}{A}$ using preimage for a given $y$.
	\end{itemize}
\end{itemize}

There are two HTLC transactions per HTLC with preimage $y$ that depend on $\commitmentTX{n}{A}$: An HTLC timeout transaction $\timeoutTX{ny}{A}$ and an HTLC success transaction $\successTX{ny}{A}$.
The inputs for the HTLC transactions are the outputs indicated above for the commitment transaction $\commitmentTX{n}{A}$ and need to be signed by Alice and Bob.
Both HTLC transactions held by Alice have one output that is spendable
\begin{itemize}
    \item[]
    \begin{itemize}
        \item by Alice after delay $\deltaTxComm$, or
        \item by Bob using Alice's revocation key for state $n$.
    \end{itemize}
\end{itemize}

\begin{protocol}[tp]
  \caption{Payment from Alice $= I_0$ to Bob $= I_N$ over intermediaries $I_{1, ..., N-1}$}
  \label[prot]{protocol:overviewPaymentOverIntermediaries}
  \textit{Inputs.} $T_\mathrm{now}$ is the current time.
  \\
  \textit{Goal.} Alice has paid Bob $c$ coins.
  \begin{algorithmic}[1]
    \State Bob draws a random secret $x$ and calculates $y = H(x)$ using a secure hash function $H$. % shorten: drop 'secret'
    \State Bob sends $y$ to Alice. %(\textsc{invoice})
    \For{$i = 0$ to $N-1$}
      \State $I_i$ runs \cref{protocol:overviewUpdateChannel} with $I_{i+1}$ with parameters $(\mathtt{add}, y, c, T_\mathrm{now} + (N - i) \cdot \deltaForward)$ to add the HTLC.
    \EndFor
    \For{$i = N$ to $1$}
      \State $I_i$ runs \cref{protocol:overviewUpdateChannel} with $I_{i-1}$ with parameters $(\mathtt{redeem}, x)$ to redeem and remove the HTLC.
    \EndFor
  \end{algorithmic}
\end{protocol}

The protocol for a payment from Alice to Bob through the PCN is described in \cref{protocol:overviewPaymentOverIntermediaries}.
The payment process starts by Bob drawing a random secret value $x$ and calculating $y = H(x)$ using a cryptographically secure hash function $H$.
Bob sends $y$ to Alice who determines a payment path between peers in the PCN.
We denominate the length of the path as $N \geq 1$ and the peers on the path as $I_i, i \in \{0,...,N\}$ where $I_0$ is Alice and $I_N$ is Bob.
Starting with Alice, every peer on the path adds the HTLC to its channel with the next peer on the path using \cref{protocol:overviewUpdateChannel}.
After $I_{N-1}$ has added the HTLC to its channel with Bob, Bob redeems the payment by sending $x$ to $I_{N-1}$ and updating the channel between Bob and $I_{N-1}$ to remove the HTLC.
Going the path backwards, the HTLCs are redeemed until Alice has paid $I_1$.
The timeouts of the HTLCs on the path from Alice to Bob are of decreasing length with difference $\deltaForward$ so that when the payment is redeemed, every peer has enough time to receive the value for $x$ and forward it to the previous peer on the path.
In case a commitment transaction containing an HTLC is sent to the first layer $\mathcal{L}$, the affected user synchrony property ensures that the other peer in the payment channel sees $x$ within at most $\deltaSync$.
Because a transaction sent to the first layer will be confirmed within $\deltaConfirm$, the difference $\deltaForward$ between the HTLC timeouts of consecutive channels on one path must be greater than $\deltaSync + \deltaConfirm$.

\begin{protocol}[tp]
  \caption{Negotiate new commitment transaction for payment channel between Alice and Bob}
  \label[prot]{protocol:overviewUpdateChannel}
  \textit{Inputs.} Alice receives $(\mathtt{add}, y, c, \timeTxHTLC{y})$ or $(\mathtt{redeem}, x)$. Alice and Bob have commitment transactions for the current state number $n$. Let $m = n+1$, $o = m+1 = n+2$.
  \\
  \textit{Goal.} Alice and Bob have their version of the new commitment transaction $\commitmentTX{m}{A}$ and $\commitmentTX{m}{B}$ signed by the other party. The old commitment transactions are revoked (i.e., their revocation keys have been shared with the other party). % shorten to save space
  \begin{algorithmic}[1]
    \If{Alice's input is $(\mathtt{add}, y, c, \timeTxHTLC{y})$}
    	\State Alice creates a new commitment transaction $\commitmentTX{m}{B}$ for state number $m$ by reducing her balance by $c$ and adding an outgoing HTLC output $(c, y, \timeTxHTLC{y})$.
    	\State Alice sends $(\mathtt{add}, y, c, \timeTxHTLC{y})$ to Bob.
    	\State Bob accordingly creates $\commitmentTX{m}{A}$ by reducing Alice's balance by $c$ and adding an incoming HTLC output $(c, y, \timeTxHTLC{y})$.
	\ElsIf{Alice's input is $(\mathtt{redeem}, x)$}
		\State Alice creates a new commitment transaction $\commitmentTX{m}{B}$ for state number $m$ removing the incoming HTLC output $(c, y, \timeTxHTLC{y})$ for $y = H(x)$ and adding $c$ to her balance.
    	\State Alice sends $(\mathtt{redeem}, x)$ to Bob.
    	\State Bob accordingly creates $\commitmentTX{m}{A}$ by increasing Alice's balance by $c$ and removing the outgoing HTLC output $(c, y, \timeTxHTLC{y})$ for $y = H(x)$.
	\EndIf
	\State Alice creates the timeout transaction $\timeoutTX{ny}{A}$ and success transaction $\successTX{ny}{A}$ for every HTLC % for every active HTLC
    \State Alice signs $\commitmentTX{m}{B}$ and all $\timeoutTX{ny}{A}$ and $\successTX{ny}{A}$ and sends the signatures to Bob. %(\textsc{commitment signed})
    \State Bob verifies that the signatures are valid.
    \State Bob creates $\commitmentTX{m}{A}$ and the $\timeoutTX{ny}{B}$ and $\successTX{ny}{B}$ accordingly and sends his signature of $\commitmentTX{m}{A}$ and all $\timeoutTX{ny}{B}$ and $\successTX{ny}{B}$ to Alice. %(\textsc{commitment signed})
    \State Alice verifies that the signature is valid for $\commitmentTX{m}{A}$.
    \State Alice creates a new key pair with the secret key $s_{\mathrm{R}o\mathrm{A}}$ and the public key $p_\mathrm{RoA}$.
    \State Alice sends $s_{\mathrm{R}n\mathrm{A}}$ and $p_\mathrm{RoA}$ to Bob. %(\textsc{revoke and ack})
    \State Bob creates a new key pair with the secret key $s_{\mathrm{R}o\mathrm{B}}$ and the public key $p_{\mathrm{R}o\mathrm{B}}$.
    \State Bob sends $s_{\mathrm{R}n\mathrm{B}}$ and $p_{\mathrm{R}o\mathrm{B}}$ to Alice. %(\textsc{revoke and ack})
  \end{algorithmic}
\end{protocol}

The payment channel between Alice and Bob can be closed by either of them by sending their latest commitment transaction to the first layer $\mathcal{L}$ (see \cref{protocol:overviewCloseChannel}).
Because of the liveness property of the first layer, the commitment transaction will be confirmed within $\deltaConfirm$.

\paragraph{Differences to the Lightning Network specification}

With the presented protocol, we tried to stay close to how the protocol for the Lightning Network is specified, however, we made the following simplifications.
The Lightning Network specification uses different keys to sign HTLCs and commitment transactions to allow for separating keys in cold and hot storage\footnote{\scriptsize{\url{https://github.com/lightningnetwork/lightning-rfc/blob/master/02-peer-protocol.md\#rationale}}}. % footnote can be commented out to save space
Instead, in our protocol, Alice uses $p_\mathrm{A}$ for commitment transactions and HTLCs.
Also, for generating a revocation key for Bob, the Lightning Network uses a construction that uses a long term secret from Bob and the per-commitment secrets from Alice to produce one secret key to sign with.
Instead, we defined the protocol to use signatures by Bob's public key and using a revocation key pair per commitment transaction.
In our protocol definition, we left out how Alice finds a path for the HTLCs for a payment to Bob.
This is conducted using source routing on continuously shared topology information about the network in the Lightning Network.
Furthermore, we defined the forwarding timeout delta $\deltaForward$ as a global value.
The Lightning Network instead uses values that are chosen by the users.
Lastly, the closing of channels can be cooperative in the Lightning Network specification to allow both parties to access their funds immediately.
We argue that all these modifications are not critical for the security of the protocol but they simplify the protocol to make it easier to follow our following argumentation that shows that the properties of the \modelName{} suffice to implement a secure PCN.

\begin{protocol}[tp]
  \caption{Closing a payment channel between Alice and Bob}
  \label[prot]{protocol:overviewCloseChannel}
  \textit{Run when Alice wants to close the channel or when for any HTLC in the channel the difference between the current time $\timeNow$ and $\timeTxHTLC{y}$ is less than $\deltaConfirm + \deltaSync$ .}
  \\
  \textit{Goal.} The payment channel between Alice and Bob has been closed and Alice and Bob have their respective balances accessible outside the channel.
  \begin{algorithmic}[1]
    \State Alice sends her latest commitment transaction $\commitmentTX{n}{A}$ and the associated HTLC transactions to the first layer $\mathcal{L}$.
  \end{algorithmic}
\end{protocol}

\afterpage{\clearpage}

\section{Security Property for the Payment Channel Network Protocol based on the \modelName{}}
\label{sec-security}

We now show that the protocol of Section \ref{sec-protocol-pcn} fulfills the security property as defined in \cref{lma-security} when used with a first layer that implements the properties of the \modelName{}.
We start by defining the term \textit{correct balance}:

\begin{definition}
\label{def-correct-balance}
The \textbf{correct balance} of a user Alice is the sum of Alice's stable balance and the amounts of outgoing HTLCs that she has not received the secret for and the amounts of incoming HTLCs that she has received the secret for.
\end{definition}

To facilitate understanding of the following lemma we give a brief review of the most important relative and absolute timings that are used:
The first layer $\mathcal{L}$ confirms a valid transaction within $\deltaConfirm$, other users will see a confirmed transaction within $\deltaSync$, a user can spend their own commitment transaction's outputs after $\deltaTxComm$, and an HTLC with condition $y$ times  out at time $\timeTxHTLC{y}$.

\begin{lemma}
\label{lma-security}
The protocol defined in \cref{sec-protocol-pcn} fulfills the property \textbf{security}: At each point in time $\timeNow$, Alice can close the payment channel between Alice and Bob so that she has received at least her correct balance at time $T = \max(\timeTxHTLCMax, T_\mathrm{now}) + 2 \cdot \deltaConfirm + \deltaTxComm$ if she is honest and checks the first layer $\mathcal{L}$ for transactions  at least once every $\deltaUserCheck$ and $0 < \deltaUserCheck < \deltaTxComm - \deltaSync - \deltaConfirm$, where $\timeTxHTLCMax$ denotes the maximal timeout of all HTLCs in the channel. % of all active HTLCs in the channel
\end{lemma}

\Cref{lma-security} follows from the following arguments.
We first look at the case that Alice initiates the closing of the channel and then at the case that Bob closes the channel and Alice did not want to close the channel.
Let $n$ be the number of the latest state.
To close the channel, Alice sends her latest commitment transaction $\commitmentTX{n}{A}$ and the associated HTLC transactions to $\mathcal{L}$ at $\timeNow$ (see \cref{protocol:overviewCloseChannel}).
Alice has Bob's signature for $\commitmentTX{n}{A}$ because she receives Bob's signature for the initial commitment transaction $\commitmentTX{1}{A}$ during the opening of the channel ($n = 1$) and during each update of the channel ($n > 1$), Alice receives Bob's signature for the latest commitment transaction $\commitmentTX{n}{A}$ and the associated HTLC transactions.
If there are no conflicting transactions, $\commitmentTX{n}{A}$ will be confirmed within $\deltaConfirm$ according to the liveness property of $\mathcal{L}$.
The funding transaction can only be spent by Alice and Bob and thus a transaction conflicting with $\commitmentTX{n}{A}$ can only be sent to $\mathcal{L}$ by Bob until $\commitmentTX{n}{A}$ is confirmed by $\mathcal{L}$.
At time $\timeNow + \deltaConfirm$ one commitment transaction will be confirmed -- either Alice's transaction or a transaction sent by Bob.
Thus, we need to distinguish three scenarios:

\begin{figure}[tbp]
    \includegraphics[width=0.9\textwidth]{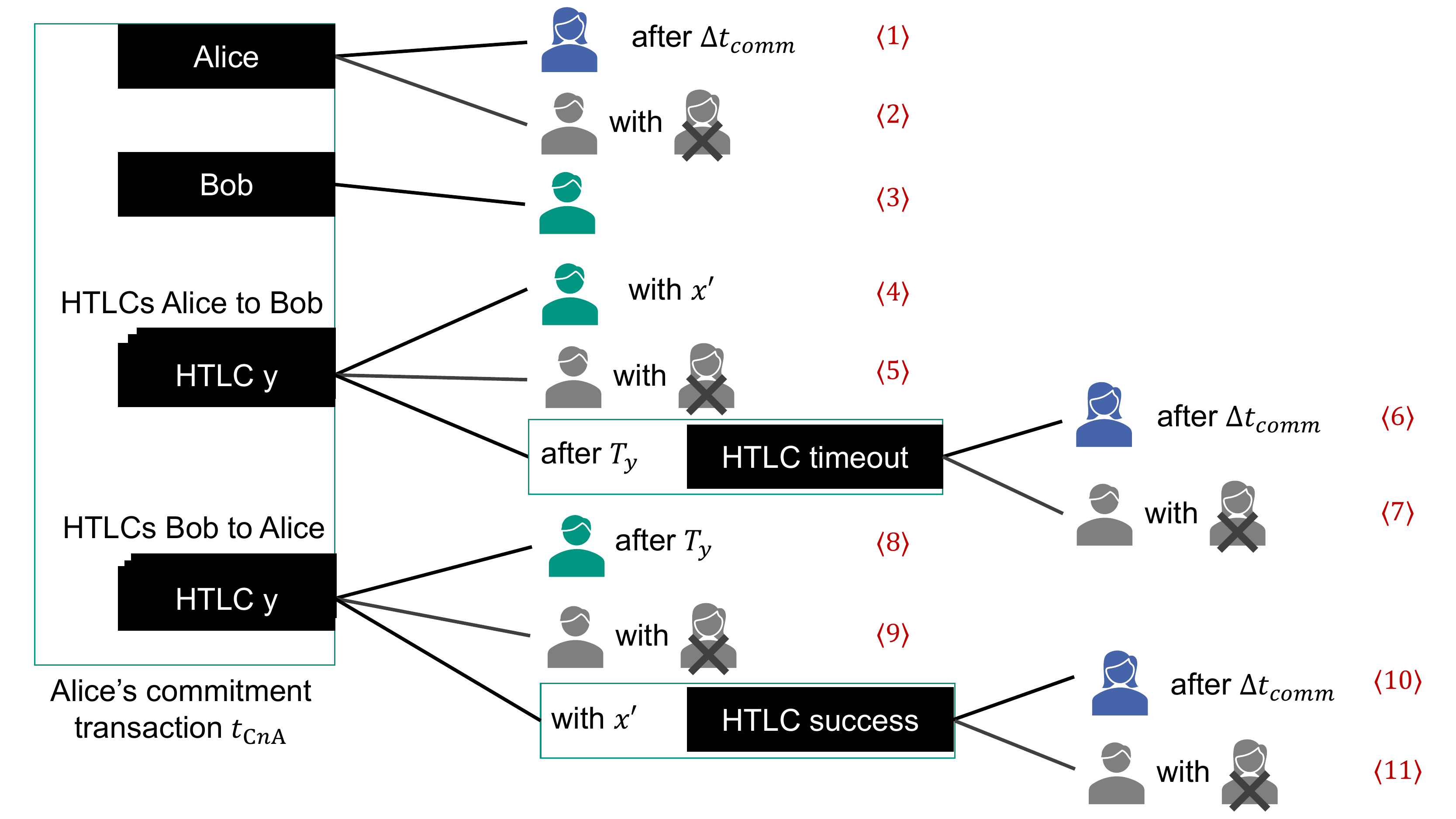}
    \caption{Ways the outputs of Alice's latest commitment transaction $\commitmentTX{n}{A}$ can be spent after the first layer has confirmed $\commitmentTX{n}{A}$.
    The blue icon indicates that Alice's secret key $s_\mathrm{A}$ is required to spend the output. Analogously, the green icon for Bob's secret key $s_\mathrm{B}$.
    The crossed-out icon stands for Alice's revocation key $s_{\mathrm{R}n\mathrm{A}}$. Paths that cannot be gone because no party can fulfill the necessary conditions on their own are grayed out. $x'$ is a preimage for $y$, i.e. $h(x') = y$.}
    \label{fig:commitment-tx-alice-on-chain}
\end{figure}

\textbf{Bob does not publish a commitment transaction:}
		The first layer $\mathcal{L}$ will confirm Alice's commitment transaction $\commitmentTX{n}{A}$ within $\deltaConfirm$ according to the liveness property.
		The outputs of $\commitmentTX{n}{A}$ are shown in \cref{fig:commitment-tx-alice-on-chain} and numbered from \figureSubnum{1} to \figureSubnum{11}.
		Alice can spend her stable balance after an additional $\deltaTxComm$ \figureSubnum{1}.
		The associated output cannot be spent by Bob \figureSubnum{2} because Bob does not have the revocation key $s_{\mathrm{R}n\mathrm{A}}$ for the latest commitment transaction.
		The output for Bob's stable balance \figureSubnum{3} does not count to Alice's correct balance.
		For the incoming HTLCs in $\commitmentTX{n}{A}$ that Alice has the secret for, she publishes the success transactions $\successTX{ny}{A}$ together with $\commitmentTX{n}{A}$ and the success transactions' outputs can be spent by Alice after $\deltaTxComm$ \figureSubnum{10}.
		Bob cannot spend the HTLC output because the associated time $\timeTxHTLC{y}$ has not come \figureSubnum{8} (else, Alice would have removed the HTLC or timely gone on-chain, see \cref{protocol:overviewCloseChannel}) and Bob does not have the revocation keys \figureSubnum{9, 11}.
		Thus, Alice can spend her stable balance and her balance of incoming HTLCs at $\timeNow + \deltaConfirm + \deltaTxComm$.
    	Her transaction to spend these outputs will be confirmed within $\deltaConfirm$.
		For the outgoing HTLCs in $\commitmentTX{n}{A}$ that Alice does not have the secret for, she can spend the output using the timeout transaction $\timeoutTX{ny}{A}$ after $\timeTxHTLC{y}$.
		The timeout transaction is confirmed by $\mathcal{L}$ and its output is spendable \figureSubnum{6} after $\timeTxHTLC{y} + \deltaConfirm + \deltaTxComm$.
    	If Bob spends the HTLC using the preimage \figureSubnum{4}, Alice receives the preimage because of the affected user synchrony property of $\mathcal{L}$ because Alice is a potential sender of the HTLC output, so the HTLC's amount is not taken into account for this channel's correct balance.
		Thus, Alice has received her correct balance after at most $\max(\timeNow, \timeTxHTLCMax) + 2 \cdot \deltaConfirm + \deltaTxComm = T$.

\begin{figure}[tbp]
    \includegraphics[width=0.9\textwidth]{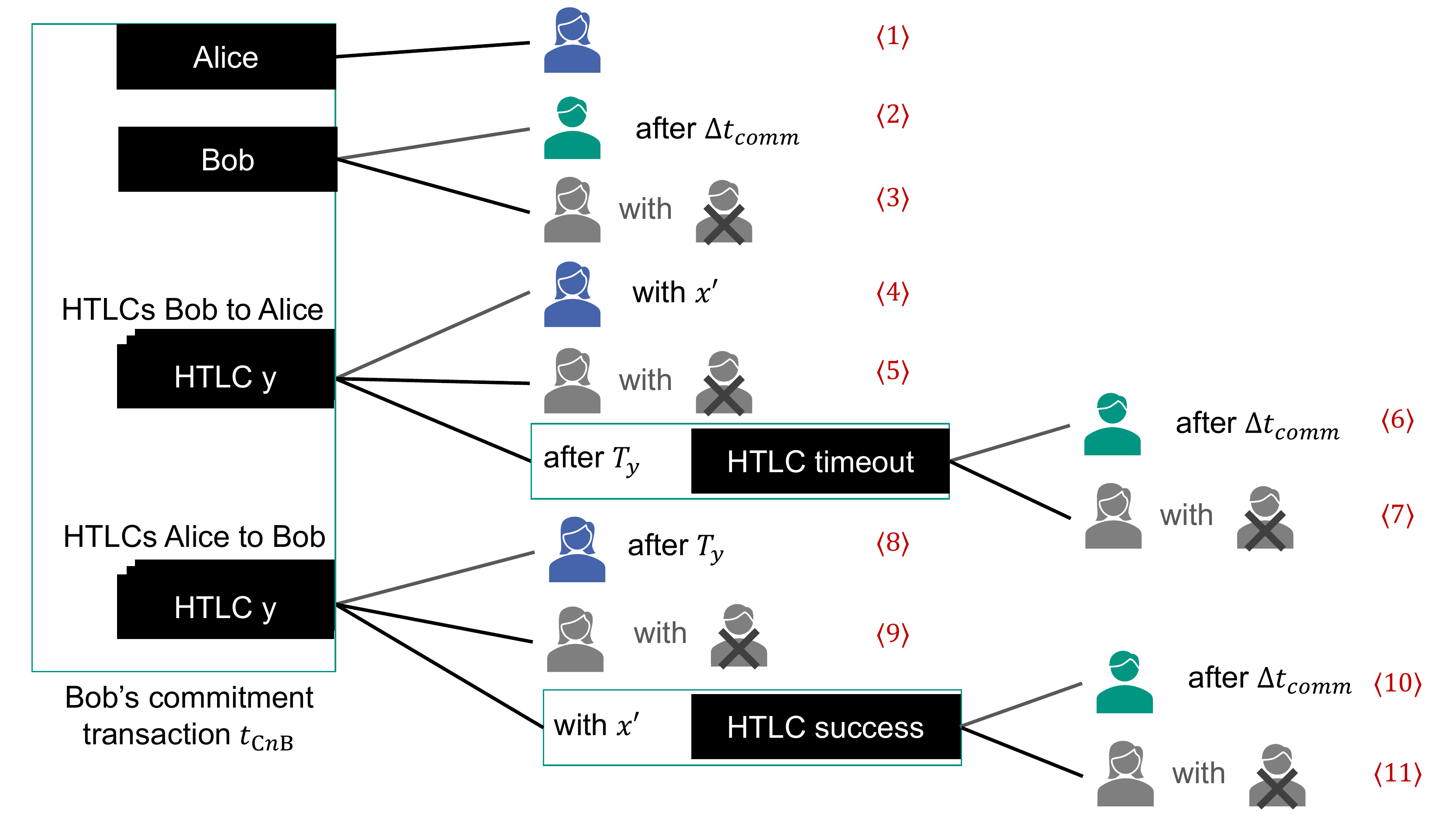}
    \caption{Ways the outputs of Bob's latest commitment transaction $\commitmentTX{n}{B}$ can be spent after the first layer has confirmed $\commitmentTX{n}{B}$.
    The blue icon indicates that Alice's secret key $s_\mathrm{A}$ is required to spend the output. Analogously, the green icon for Bob's secret key $s_\mathrm{B}$.
    The crossed-out icon stands for Bob's revocation key $s_{\mathrm{R}o\mathrm{B}}$. Paths that cannot be gone because no party can fulfill the necessary conditions on their own are grayed out. $x'$ is a preimage for $y$, i.e. $h(x') = y$.}
    \label{fig:commitment-tx-bob-on-chain}
\end{figure}
		
\textbf{Bob has published his latest commitment transaction $\commitmentTX{n}{B}$} (see \cref{fig:commitment-tx-bob-on-chain}):
		In this case, the affected user synchrony property of the first layer asserts that Alice can see Bob's commitment transaction $\commitmentTX{n}{B}$ because she is a potential sender.
		Alice can instantly spend her stable balance in the channel \figureSubnum{1} once she sees the transaction $\commitmentTX{n}{B}$ at time $\timeNow + \deltaConfirm + \deltaSync$.
		Alice's transaction spending her stable balance will be confirmed within $\deltaConfirm$.
		For each outgoing HTLC, Alice can spend the HTLC output after $\timeTxHTLC{y}$ \figureSubnum{8}.
		If Bob spends the HTLC output using $\successTX{ny}{B}$ by providing a preimage for the given $y$ \figureSubnum{10}, Alice receives the preimage because of the affected user synchrony property of $\mathcal{L}$.
		For each incoming HTLC, Alice must publish a transaction to redeem the HTLC \figureSubnum{4} if she has the preimage for the given $y$ which takes $\deltaConfirm$ to be confirmed. % must because Bob could publish a timeout later
		Thus, Alice has received her correct balance within $\max(\timeNow + \deltaConfirm + \deltaSync, \timeTxHTLCMax) + \deltaConfirm \leq T$.

\begin{figure}[tbp]
    \includegraphics[width=0.9\textwidth]{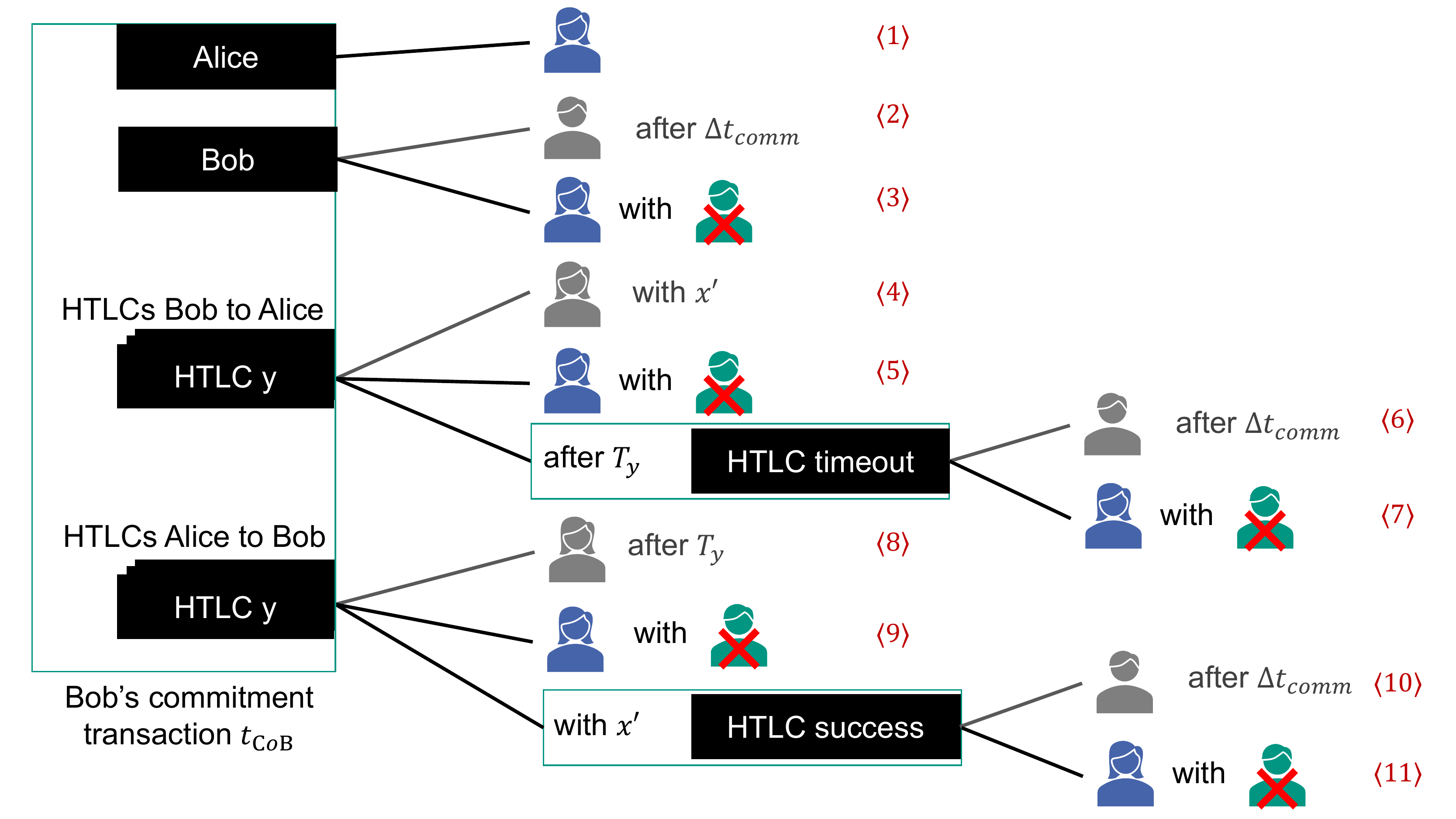}
    \caption{Ways the outputs of Bob's outdated commitment transaction $\commitmentTX{o}{B}$ can be spent after the first layer has confirmed $\commitmentTX{o}{B}$.
    The blue icon indicates that Alice's secret key $s_\mathrm{A}$ is required to spend the output. Analogously, the green icon for Bob's secret key $s_\mathrm{B}$.
    The crossed-out green icon stands for Bob's revocation key $s_{\mathrm{R}o\mathrm{B}}$. Paths that cannot be gone because no party can fulfill the necessary conditions on their own are grayed out. $x'$ is a preimage for $y$, i.e. $h(x') = y$.}
    \label{fig:commitment-tx-bob-outdated}
\end{figure}
		
\textbf{Bob has published an outdated commitment transaction} (see \cref{fig:commitment-tx-bob-outdated}):
		For each update to state number $i+1$, Alice receives Bob's revocation key $s_{\mathrm{R}i\mathrm{B}}$.
		Say, Bob has published an outdated commitment transaction $\commitmentTX{o}{B}$ with $o < n$.
		Alice can see the transaction after $\timeNow + \deltaConfirm + \deltaSync$.
		Bob can only spend his stable output of $\commitmentTX{o}{B}$ after an additional $\deltaTxComm$ \figureSubnum{2}.
		In case Bob publishes an HTLC success \figureSubnum{10,11} or timeout transaction \figureSubnum{10,11}, Alice also sees Bob's HTLC transaction because of the affected user synchrony property of $\mathcal{L}$ and Bob can only spend its output after $\deltaTxComm$ \figureSubnum{6,10}.
		Thus, Alice must use her key $s_\mathrm{A}$ and Bob's revocation key $s_{\mathrm{R}o\mathrm{B}}$ to create a revocation transaction that spends the whole balance in the channel (output for Alice \figureSubnum{1}, output for Bob \figureSubnum{3}, and all HTLC outputs \figureSubnum{7,11}).
		After $\deltaConfirm$ this revocation transaction has been confirmed.
		Because $\deltaConfirm < \deltaTxComm$ (see \cref{lma-security}), Bob cannot have spent his outputs before Alice.
		Thus, Alice has received her correct balance within $\timeNow + 2 \cdot \deltaConfirm + \deltaSync \leq T$.
		
In case Bob closes the channel by sending $\commitmentTX{i}{B}, i \leq n$ to $\mathcal{L}$ at $\timeNow$ and Alice did not want to close the channel, too, Alice receives $\commitmentTX{i}{B}$ from $\mathcal{L}$ within $T_\mathrm{recvA} = \timeNow + \deltaConfirm + \deltaSync + \deltaUserCheck$ because Alice is an affected user of the transaction and the affected user synchrony property of $\mathcal{L}$ asserts that she can see the transaction within $\deltaSync$ and Alice checks the first layer $\mathcal{L}$ at least every $\deltaUserCheck$ for new transactions.
Bob's stable output and HTLC transaction outputs cannot be spent by him until $T_\mathrm{spendB} = \timeNow + \deltaConfirm + \deltaTxComm$.

If $i < n$, Alice must use her key $s_\mathrm{A}$ and Bob's revocation key $s_{\mathrm{R}i\mathrm{B}}$ to spend the whole balance in the channel using a revocation transaction when she sees Bob's transaction at time $T_\mathrm{recvA}$.
This revocation transaction will be confirmed by the first layer after $T_\mathrm{recvA} + \deltaConfirm$.
Bob cannot have spent his outputs at this time because $T_\mathrm{recvA} + \deltaConfirm = \timeNow + \deltaConfirm + \deltaSync + \deltaUserCheck + \deltaConfirm < \timeNow + \deltaConfirm + \deltaTxComm = T_\mathrm{spendB}$ because $\deltaUserCheck < \deltaTxComm - \deltaSync - \deltaConfirm \implies \deltaConfirm + \deltaSync + \deltaUserCheck < \deltaTxComm$.
Thus, Alice has received her correct balance after $\timeNow + \deltaConfirm + \deltaSync + \deltaUserCheck + \deltaConfirm < \timeNow + \deltaConfirm + \deltaTxComm \leq T$.

If $i = n$, Alice can immediately spend her stable balance when she receives $\commitmentTX{i}{B}$ at time $T_\mathrm{recvA}$.
Alice reacts analogously to the case that Bob has published his latest commitment transaction (see \cref{fig:commitment-tx-bob-on-chain}) but the times are postponed by $\deltaUserCheck$.
It follows from the same argumentation that Alice has received her correct balance within $\max(\timeNow + \deltaConfirm + \deltaSync + \deltaUserCheck, \timeTxHTLCMax) + \deltaConfirm \leq T$.

\section{Instances and Options of the \modelName}
\label{sec-instances}

The \modelName{} describes an ideal first layer that guarantees the properties required by a PCN.
In this section, we show that, under certain assumptions, a blockchain instantiates such a first layer.
We also sketch the idea of an instance of a first layer using a bank or a network of banks and provide a comparative exploration of design options.

{\bf Using a Blockchain.} 
Garay et al. show in various works (e.g., \cite{garay_bitcoin_2015,garay_full_2020}) that the Bitcoin protocol satisfies \textit{consistency} and \textit{liveness} with high probability under the assumption of a bounded-delay network model and an honest majority of computing power \cite{garay_full_2020}.
In comparison to our definition in \cref{sec-model-first-layer}, their definition of liveness assumes that a transaction is provided to all honest parties.
This is implemented in Bitcoin by flooding transactions in the peer-to-peer network.
The definition of consistency used in \cite{garay_full_2020} implies our definition of persistence and affected user synchrony.
It is even stronger and implies \textit{synchrony} for all honest peers, i.e. the first layer $\mathcal{L}$ makes a transaction $t$ and the confirmation visible to all honest peers.
Assuming an honest majority of computing power and using a bounded-delay network model, the results of Garay et al.\ show that a blockchain similar to Bitcoin instantiates the \modelName{} with high probability.

As forks can occur in a blockchain, reorganizations of blocks can affect the persistence property.
The more descendants a block has, the smaller is the probability for the block to be invalidated by the longest chain rule.
Increasing $\deltaConfirm$ increases the probability that the persistence property is fulfilled.

Note that for a blockchain, liveness is not guaranteed because the blocksize is limited and there can be times during which the blockchain is congested so that users have to compete for publishing their transactions on the blockchain.
A user might have to pay higher transaction fees to include a transaction or might not be able to include a transaction at all.
For building a PCN, it is assumed that the congestion does not last longer than $\deltaConfirm$ or that a user is willing to pay a sufficiently high transaction fee.
Also, a reliable network link is required for reliable exchange of messages to satisfy liveness.
These assumptions need to be met to instantiate a first layer that allows for a secure PCN as second layer.
Recent work \cite{khabbazian_timelocked_2020,tsabary_mad-htlc_2020,harris_flood_2020} has shown attacks against payment channels that attack the liveness of a blockchain, e.g. by bribing miners to censor transactions.
These works show the importance of considering the properties of the first layer when building second layer architectures.

{\bf Using a Single or Multiple Banks.} Having an abstract model of the first layer allows for developing architectures that instantiate a first layer without a blockchain.
For example, a network of banks can be used to instantiate a first layer under the assumption of trust in the banks.
We assume the common features of banks as described in the following but the bank does not necessarily need to be a classical bank and can also be a payment service provider.

A contemporary bank offers to their consumers an interface that implements liveness, transaction validity, and persistence.
The usual visibility for a transaction matches the visibility required by affected user synchrony:
A bank makes a transaction visible (only) to the potential senders and receivers of transferred funds.
A transaction has multiple potential senders if it is sent from a joint account.

Using banks as first layer, their customers could perform transactions using a PCN.
While this requires trust into the bank to implement the \modelName{}, the transactions are hidden from the bank which improves privacy because the bank gains less information.
So the PCN could be used for decentralized digital cash.

The interface that banks offer their customers is an account model and not a UTXO based transaction model.
This does not constitute a contradiction to the \modelName{} of a first layer.
On the one hand, banks could implement a UTXO model in the backend and show the customer an interface matching the account model by grouping UTXOs of one customer together to generate a view of a virtual account.
On the other hand, the model of the first layer can also be transformed to use an account model and smart contracts, as implemented by the Raiden Network on the Ethereum blockchain.

{\bf Comparative Exploration of Design Options.} Having seen that there can be different ways to instantiate the \modelName{} for a first layer for a PCN, we now discuss differences between using a public permissionless blockchain such as Bitcoin and trusted banks to implement a first layer.
These differences show that using other instances of the \modelName{} instead of a blockchain creates new design options.

\textit{Trust}
A basic difference between a public permissionless blockchain and trusted banks are the trust assumptions.
While it is clear that banks have to be fully trusted -- the trust might be backed by trusting the legal framework to provide justice in case of fraud by the bank --, the trust required into a public permissionless blockchain is more distributed.
Users running payment channels on a blockchain such as Bitcoin have to trust that Bitcoin implements the properties of the \modelName{}.
Especially for liveness, it needs to be assumed that enough miners do not censor transactions (see above) and honest miners have the largest share of computation power.

\textit{Privacy}
Whether PCNs in general improve privacy because the visibility of transactions is restricted is unclear and recent research \cite{kappos_empirical_2020,rohrer_counting_2020,romiti_cross-layer_2020} shows that privacy is not improved categorically.
Also, the privacy properties of the first layer are crucial for privacy in the PCN.
On a public permissionless blockchain, users are identified by pseudonyms.
Each user can generate as many pseudonyms as they wish which makes tracking users more difficult despite all transactions being public.
Contrary to blockchains, a bank is required to implement methods for customer identification.
This reduces privacy for the users because the bank learns about their transactions.
However, it allows the bank to implement access control on the transactions and to make a transaction only visible to the affected users of the transaction which, in turn, improves privacy because unaffected users do not learn about the transactions.
By facilitating tracing of money laundering, customer identification can be a way to increase chances of mainstream adoption of a digital payment system.

\textit{Liquidity}
Payment channels require users to deposit funds by locking them on the first layer while the channel is open.
On a public permissionless blockchain, users can only use coins in their channel that they own on the blockchain.
Using banks to implement a first layer allows for letting users open channels using credit they receive from their bank.
Similar to a credit card, a bank could give their customers credit for the second layer and demand being payed back after a set time.
This can improve the liquidity inside the network because more users are able to forward payments when they have channels with higher capacity.

\textit{Online requirement}
While the first layer needs to provide affected user synchrony to make transactions visible, the corresponding part for the user is to check the first layer regularly for new transactions to be able to react to outdated commitment transactions.
With a blockchain as a first layer, this task requires a user to stay connected and analyze new blocks.
As a blockchain is a decentralized system, a user can improve the resilience by being connected to multiple peers and implement measures against being eclipsed from the network \cite{heilman_eclipse_2015}.
If a user cannot fulfill the online requirement on their own, they can delegate this task to a watchtower.
Such proposals have been discussed by the Lightning Network's community \cite{dryja_unlinkable_2016,osuntokun_hardening_2018} and by peer-reviewed research \cite{leinweber_tee-based_2019,mccorry_pisa_2019,khabbazian_outpost_2019,avarikioti_cerberus_2020-1}.
With a centralized system of banks as first layer, the bank needs to be trusted to run a system that provides affected user synchrony.
However, the advantage in this case is that the bank knows their customers and could even contact them to inform them about relevant transactions if a timely reaction is required.
Thus, the bank could include the watchtower functionality in the first layer.

\textit{Currencies}
The two different ways for instantiating the \modelName{} also differ in the type of currencies they can support.
While a public permissionless blockchain can provide a decentralized currency, a system of banks can make traditional currencies managed by central banks available for use in PCNs.

\section{Optimization of HTLCs using a Blockchain}
\label{sec-optimizations}

The basic construction of PCNs leaves room for optimizations.
One issue is that the amount of collateral that needs to be locked for one transaction allows for balance availability attacks \cite{perez-sola_lockdown:_nodate} which lock the balance so that it cannot be used by honest parties.
Recent research \cite{miller_sprites_2017} has found ways to reduce the required collateral during one transaction over multiple hops assuming a blockchain as first layer that offers (global) synchrony instead of the reduced affected user synchrony.
This protocol makes use of a more generalized synchrony property that is implemented by blockchains but not part of our reduced model of a first layer:
Using a blockchain, a transaction can be seen by any party (synchrony) while using the \modelName{}, only the affected users of a transaction are guaranteed to see the transaction (affected user synchrony).
This possibility of making a value visible and usable for the public, is used by \cite{miller_sprites_2017}.
The Sprites protocol \cite{miller_sprites_2017} uses a (logically) central ``preimage manager'' that can be read and written to by any party of the PCN.
All channel updates on the route of a payment depend on the condition that the secret $x$ has been published before a specific timeout that is the same across all channels.
By using it as a ``global synchronization gadget'', the preimage manager is used to synchronize the timeouts so that all parts of the route timeout at the same time in contrast to increasing timeouts from the receiver's end in the Lightning Network.
The preimage manager is used to publish the secret in case a party is not responsive or misbehaves.
As all participants have the same view on the blockchain, either all updates that depend on the publication of the secret before a given timeout fail or all are valid.
For a first layer instantiated by banks, a bank or another trusted party could implement such a preimage manager.
However, this would add a central entity as dependency for the payments and thus, payments would not be decentralized anymore.
This shows that, while the \modelName{} allows for a PCN to be implemented as second layer, the stronger properties fulfilled by a blockchain enable optimizations for HTLCs.
The solution of \cite{miller_sprites_2017} can improve the payment channel protocol because they make use of the gap between these two models. % this is SoK

\section{Related Work}
\label{sec-related-work}

Kiayias and Litos provide in \cite{kiayias_composable_2020} a formalisation and security analysis of the Lightning Network using a global ledger functionality modeled in \cite{badertscher_bitcoin_2017}.
The global ledger fulfills the synchrony property mentioned above: An honest user that is connected to the required resources is being synchronized (receives the latest state) within a bounded time.
Their work is orthogonal to our work because it appears that their proof would still work with the assumptions of the \modelName{}.
We leave it for future work to provide a security proof for the Lightning Network that uses the \modelName{} instead of the global ledger functionality provided by Bitcoin.

Credit networks as proposed in \cite{fugger_money_2004} are a concept that is related to that of PCNs.
A credit network, however, does not have an underlying layer.
In a credit network, users are connected through credit links (IOUs) which represent the amount one user owes another user.
This construction requires users to trust each other to an extent that is quantified by the size of the credit link.
In a PCN, users are instead required to trust the first layer and not each other.
While credit networks work without a first layer, we model the required properties of a first layer and show different ways to implement such a first layer.

Avarikioti et al.\ propose in \cite{avarikioti_cerberus_2020-1} a protocol for payment channels that includes watchtowers that watch for outdated commitment transactions on the first layer.
They also define a security property and show that the protocol for Cerberus channels fulfills this property.
The security property we define in \cref{lma-security} is inspired by the security property used in \cite{avarikioti_cerberus_2020-1}; however, our definition does explicitly consider the timeouts and includes HTLCs which are required for payments over intermediaries.

\section{Conclusion}
\label{sec-conclusion}

For a PCN, a first layer can be used that delivers only a reduced set of properties compared to a blockchain.
We defined these properties in the \modelName{} and showed that this model suffices to implement a secure protocol for PCNs.
Furthermore, the \modelName{} can be instantiated by blockchains.
We examined how the difference between the properties that blockchains have in comparison to the \modelName{} can be used to improve payments over HTLCs and we have shown that this difference has already been used by improvements that have been proposed in previous works.
We also showed that banks can instantiate the \modelName{}.
Implementing a first layer might be a role banks play in the future.

\bibliographystyle{splncs04}
\bibliography{library}

\end{document}